# Large Anomalous and Topological Hall Effect and Nernst Effect in a Dirac Kagome Magnet Fe$_3$Ge

Chunqiang Xu[1,2], Shuvankar Gupta[2], Hengxin Tan[3], Hyeonhu Bae[3], Olajumoke Oluwatobiloba Emmanuel[2], Mingyu Xu[4], Yan Wu[5], Xiaofeng Xu[6], Pengpeng Zhang[2], Weiwei Xie[4], Binghai Yan[3], Xianglin Ke[2]

[1] School of Physical Science and Technology, Ningbo University, Ningbo 315211, China
[2] Department of Physics and Astronomy, Michigan State University, East Lansing, Michigan 48824, USA
[3] Department of Condensed Matter Physics, Weizmann Institute of Science, Rehovot, Israel
[4] Department of Chemistry, Michigan State University, East Lansing, Michigan 48824, USA
[5] Neutron Scattering Division, Oak Ridge National Laboratory, Oak Ridge, Tennessee 37831, USA
[6] Department of Applied Physics, Zhejiang University of Technology, Hangzhou 310023, China

Searching for Kagome magnets with novel magnetic and electronic properties has been attracting significant efforts recently. Here, we report magnetic, electronic and thermoelectric properties of Fe$_3$Ge single crystals with Fe atoms forming a slightly distorted Kagome lattice. We showed that Fe$_3$Ge exhibits large anomalous Hall effect and anomalous Nernst effect. The observed anomalous transverse thermoelectric conductivity $|\alpha_{xz}^A|$ reaches ~ 4.6 A m$^{-1}$ K$^{-1}$, which is larger than the conventional ferromagnets and most of topological ferromagnets reported in literature. Our first-principles calculations suggest that these exceptional transport properties are dominated by the intrinsic mechanism, which highlights the significant contribution of the Berry curvature of massive Dirac gaps in the momentum space. Additionally, a topological Hall resistivity of 0.9 $\mu\Omega$ cm and a topological Nernst coefficient of 1.2 $\mu$V/K were also observed, which are presumably ascribed to the Berry phase associated with the field-induced non-zero scalar spin chirality. These features highlight the synergic effects of the Berry phases in both momentum space and real space of Fe$_3$Ge, which renders it an excellent candidate for room-temperature thermoelectric applications based on transverse transport.

Magnetic Kagome lattice, which is composed of corner-sharing triangles of magnetic ions, gives rise to a rich variety of electronic and magnetic properties[1-11]. The complex interplay between spin, orbital, and lattice degrees of freedom in magnetic Kagome systems can lead to frustrated magnetism, topological states, and potentially novel quantum phases, etc. For instance, various compounds with a magnetic Kagome lattice have recently been reported to exhibit intriguing electronic properties, such as frustrated antiferromagnetic $Gd_2PdSi_3$ possessing a Skrymion lattice that gives rise to a giant topological Hall effect[12], ferrimagnetic $TbMn_6Sn_6$ with topological Chern magnetism[13-15], topological Dirac semimetal $Fe_3Sn_2$[16-18], ferromagnetic Weyl semimetal $Co_3Sn_2S_2$[19,20], and antiferromagnetic FeGe with charge-density wave (CDW) state[21,22], etc. In particular, magnets hosting topological Dirac fermions owing to the Kagome lattice tend to show anomalous transverse transport phenomena and have garnered intensive attention. Specifically, in ferromagnetic Kagome systems, when the Dirac cones are proximate to the Fermi level, the time-reversal symmetry breaking can induce an energy gap opening at the Dirac points. As a result, a significant Berry curvature emerges in the momentum space which can consequently generate an intrinsic anomalous Hall effect (AHE) and an anomalous Nernst effect (ANE, which refers to an induced electric voltage between two contact leads transverse to the applied thermal gradient along the longitudinal direction). For instance, a colossal ANE was observed in a topological ferromagnetic Kagome metal $UCo_{0.8}Ru_{0.2}Al$ with the Nernst effect coefficient reaching ~ a record value of 23 $\mu V/K$ [23]. Generally, the ANE is sensitive to the Berry curvature of electron bands near the Fermi level and is approximately proportional to the energy derivative of electric Hall conductivities at the Fermi level according to the Mott relationship[24-28].

Despite the novel properties of Kagome magnets hosting Dirac fermions, compounds with an ideal magnetic Kagome lattice that are suitable for room-temperature spintronic and thermoelectric applications remain scarce. This is due to the intrinsic coordination requirements of the magnetic Kagome lattice, which renders interatomic interactions highly sensitive to structural stability, with even minor perturbations capable of compromising its structural integrity. Very recently, by utilizing angle-resolved photoemission spectroscopy and density functional theory calculations, it was proposed that Fe$_3$Ge, a ferromagnetic compound with an ordering temperature of ~ 660 K and a slightly distorted Kagome lattice, exhibits Dirac fermions[29]. And the topological nature of electronic properties has been supported by the observation of AHE[30,31] and topological Hall effect[31] in polycrystalline Fe$_3$Ge. Compared to polycrystalline samples, the absence of grain boundaries in single crystals offers an opportunity to investigate the intrinsic electronic and thermal transport properties. In addition, single crystals also enable the study of anisotropic electronic and magnetic properties, which are intimately correlated to the crystalline anisotropy. Therefore, it is desirable to investigate the electronic and thermoelectric properties of Fe$_3$Ge single crystals, which, in conjunction with theoretical calculations, allows for a more comprehensive understanding of the topological properties of this system.

In this paper, we report the key topological characteristics of electronic and thermoelectric transport properties in Fe$_3$Ge. We show that Fe$_3$Ge exhibits giant AHE and ANE. The observed anomalous thermoelectric transverse conductivity reaches ~ 4.6 A m$^{-1}$ K$^{-1}$. Our first-principles calculations find several Dirac band crossings near the charge neutral point and reproduce qualitatively the observed AHE and ANE, indicating the intrinsic origin of the anomalous transport properties in Fe$_3$Ge and highlighting the importance of its topological electronic structure. In addition, a topological Hall resistivity with a value of ~ 0.9 $\mu\Omega$ cm and topological Nernst coefficient of ~ 1.2 $\mu$V/K were observed at 320 K, which is ascribable to the Berry curvature associated with the field-induced non-zero scalar spin chirality. These features, together with its high magnetic ordering temperature, place Fe$_3$Ge an ideal candidate for room-temperature thermoelectric applications based on Nernst effect.

**Crystal structure and magnetic properties** Figure 1A shows a photo image of Fe$_3$Ge rod-like single crystals grown using a Tin flux method. Fe$_3$Ge crystallizes in hexagonal D0$_{19}$ type structure with the space group *P6$_3$/mmc* (No.194). The magnetic Fe ions form a Kagome plane with a non-magnetic Ge ion occupying at the center of each hexagon. Interestingly, as also detailed in the Supplemental Information, in Fe$_3$Ge the Fe atoms do not occupy the special positions at $x = 1/6$ and $y = 1/3$. As illustrated in Figure 1B, while six Fe ions still form a regular hexagon, the Ge ion is slightly off-center of the hexagon, which results in γ ≠ β ≠ α = 60°. Note that within each hexagon the two adjacent Fe-Fe bond lengths are 2.672 Å and 2.497 Å, indicating a slightly distorted magnetic Kagome lattice. Each Kagome plane is alternately stacked along the *c*-axis as shown in Figure 1C. Detailed single crystal X-ray characterizations, including four-circle single-crystal X-ray diffraction and the comprehensive structural analysis, as well as Laue diffraction measurements, are described in the Supplemental Information.[32]

Figure 1D shows the temperature dependence of magnetic susceptibility with the magnetic field applied along the *c*-axis and the *ab* plane respectively. Note that there is no difference in magnetic susceptibility between field-cooled and zero-field-cooled measurements. The system becomes ferromagnetically order below $T_c$ ~ 660 K with the magnetic easy-axis aligned along the *c*-axis[33,34]. This is followed by a spin reorientation transition at $T_{SR}$ ~ 385 K with the magnetic moment direction changed from the *c*-axis to the *ab* plane, which is evidenced by the sharp drop in magnetic susceptibility measured along the *c*-axis. The spin reorientation transition is supported by single-crystal neutron diffraction measurements. Figure 1E plots the temperature dependence of integrated neutron scattering intensity of (1 0 0) and (1 0 1) Bragg reflections, both of which include both nuclear and magnetic scattering intensity contribution. Note that neutrons couple to the magnetic moment perpendicular to the momentum transfer vector **q**. Thus, the decrease in the scattering intensity of (1 0 0) Bragg peak and the increase of (1 0 1) Bragg peak intensity below $T_{SR}$ indicates the spin-orientation. The magnetic structures obtained by refining a series of Bragg peaks measured at 410 K and 360 K are illustrated in the insets of Figure 1E. At 410 K spins align along the *c*-axis, while at 360 K there is a large magnetic moment component (~ 1.93 $u_B$/Fe) lying within the *ab* plane and ~ 0.45 $u_B$/Fe along the *c*-axis, respectively. The refined total magnetic moment is ~ 1.98 $u_B$/Fe, which agrees with the value obtained from isothermal magnetization *M(H)* measurements and from previous neutron powder diffraction measurements[33]. Figure 1F presents a typical *M(H)* measured with *H // c* and *H // ab* at 300 K.

**Resistivity and Hall conductivity** Figure 2A shows the schematic diagram of the longitudinal resistivity, Hall effect, and thermoelectric measurements. For the longitudinal resistivity $\rho_{zz}$, as an example, both the electric current and longitudinal voltage are along the z-axis. Figure 2B presents the temperature dependence of longitudinal resistivity $\rho_{xx}$ and $\rho_{zz}$. Both $\rho_{xx}$ and $\rho_{zz}$ decrease with the decreasing temperature, with a large residual resistivity ratio (RRR) of ~ 20 and ~ 40 for in-plane and out-of-pane measurements, respectively, affirming the high quality of Fe$_3$Ge single crystals. Unlike most other Kagome materials, the value of the in-plane resistivity $\rho_{xx}$ is about 3 ~ 4 times of the out-of-plane resistivity $\rho_{zz}$ over the entire temperature range measured. Such an anisotropic electronic transport property presumably arises from the stronger orbital hybridization between neighboring 3$d$ electrons of Fe ions due to shorter Fe-Fe bond length along the out-of-plane direction compared to that within the $ab$ plane. Note that similar anisotropic character between $\rho_{xx}$ and $\rho_{zz}$ in Fe$_3$Ge has also been reported in Ref. [29] recently.

Figure 2C shows the magnetic field dependence of Hall resistivity $-\rho_{xz}$ measured at various temperatures, where an electrical current is applied along the longitudinal direction ($c(z)$-axis) and the magnetic field is along the perpendicular direction ($ab$-plane). To eliminate the Hall resistivity contribution from the voltage probe misalignment, all data measured at different temperatures were processed by $\rho_{xz} = (\rho_{xz}(+\mu_0 H) - \rho_{xz}(-\mu_0 H))/2$ since the Hall signal is antisymmetric relative to magnetic field. As expected for ferromagnetic metals, the Hall resistivity $\rho_{xz}$ rapidly increases with the magnetic field and then saturates when the saturation magnetization is nearly reached, showing the AHE. Compared to the dominant AHE, the small slope at high magnetic field contributed by the ordinary Hall effect indicates that the ordinary Hall effect is negligible. The AHE value at 320 K is ~ $2\mu\Omega\ cm$, which is comparable with those reported in some well-known magnetic Kagome materials, e.g., Co$_3$Sn$_2$S$_2$ and Fe$_3$Sn$_2$[16,19].

To figure out the mechanism of the observed AHE in Fe$_3$Ge, we plot the scaling of $|\rho_{xz}|$ versus $\rho_{zz}\rho_{xx}$. As shown in Figure 2D, the good linear relation between $|\rho_{xz}|$ and $\rho_{zz}\rho_{xx}$ suggests that the extrinsic skew-scattering which generally gives $|\rho_{xz}| \propto (\rho_{zz}\rho_{xx})^{1/2}$ can be ruled out and the intrinsic Berry phase or extrinsic side-jump mechanisms are dominant in Fe$_3$Ge. In Figure 2E, we plot the magnetic field dependence of Hall conductivity $\sigma_{xz}$, which is defined as $\sigma_{xz} = \frac{-\rho_{xz}}{\rho_{xx}\rho_{zz}-\rho_{xz}\rho_{zx}}$, where $\rho_{zx} = -\rho_{xz}$. Because $\rho_{xz}$ is much smaller than both $\rho_{xx}$ and $\rho_{zz}$ as shown in Figure 2B and 2C, $\sigma_{xz}$ thus can be approximated as $\sigma_{xz} \approx \frac{-\rho_{xz}}{\rho_{xx}\rho_{zz}}$. Theoretically, the extrinsic side-jump contribution of $\sigma_{xz}^A$ has been shown to be on the order of $(e^2/(ha))(\varepsilon_{SO}/E_F)$, where $e$ is electronic charge, $h$ is the Plank constant, $a$ is the lattice parameter, $\varepsilon_{SO}$ is the spin-orbit coupling (SOC), the $E_F$ is the Fermi energy[35]. For metallic ferromagnets, $\varepsilon_{SO}/E_F$ is generally smaller than $10^{-2}$. With $a \sim 4 - 5$ Å, the estimated extrinsic side-jump contribution is $\sim 2$ ($\Omega^{-1}cm^{-1}$), much smaller than the total $\sigma_{xz}^A$. This suggests that the AHE in Fe₃Ge is dominated by an intrinsic mechanism, which is supported by the theoretical calculation as to be discussed later. The temperature dependent $\sigma_{xz}^A$ is presented in Figure 2F. With the increase of temperature, $\sigma_{xz}^A$ first increases and then decreases, with $|\sigma_{xz}^A| \sim 350$ ($\Omega^{-1}cm^{-1}$) at 60 K and a maximum $|\sigma_{xz}^A| \sim 550$ ($\Omega^{-1}cm^{-1}$) at 220 K. The large anomalous Hall conductivity observed in Fe₃Ge is comparable to, or exceeds, those reported for other Kagome magnets such as Mn₃Sn[36], Mn₃Ge[3], Fe₃Sn₂ [16], TbMn₆Sn₆ [13,14], and Co₃Sn₂S₂ [19]. As to be discussed later, this highlights the strong Berry curvature contributions arising from the non-trivial topology of the electronic bands in Fe₃Ge. Note that during our manuscript preparation we noticed a very recent preprint[37] reporting the observation of a large anomalous Hall effect on Fe₃Ge single crystals, which however did not consider the anisotropic electronic transport between $\rho_{xx}$ and $\rho_{zz}$ in the data processing.

**Seebeck and Nernst effect** We investigated the thermoelectric transport properties of Fe₃Ge with the thermal gradient applied along the *z*-axis and the magnetic field along the *y*-axis, as illustrated by the schematics sketched in Figure 2A. Figure 3A presents the temperature dependence of Seebeck coefficient $S_{zz}$ measured at zero applied magnetic field. The negative value of $S_{zz}$ over the whole measurement temperature range indicates that the n-type (electron) charge carriers are dominant in Fe₃Ge. Figure 3B shows the magnetic field dependent Nernst signal $S_{xz}$ measured at various temperatures. Similar to the Hall resistivity $\rho_{xz}$ shown in Figure 2C, the Nernst signal $S_{xz}$ is dominated by the anomalous contribution. The normal component of $S_{xz}$ in the high field regime is negligible, as implied by the nearly field-independence of $S_{xz}$. It is noted that $S_{xz}$ is $\sim 3.2$ μV/K at 320 K, which is much larger than those of isostructural antiferromagnets Mn₃X and conventional ferromagnets[2,38-40]. To better understand the thermoelectric properties of Fe3Ge, in Figure 3C we present the calculated transverse thermoelectric conductivity $\alpha_{xz}$ at various temperatures using $\alpha_{xz} = \frac{S_{xz}\rho_{zz} - \rho_{xz}S_{zz}}{\rho_{xx}\rho_{zz}}$ with $\rho_{xx}$, $\rho_{zz}$, $\rho_{xz}$, $S_{zz}$, $S_{xz}$ being the measured quantities[41]. It is clearly seen that $\alpha_{xz}(H)$ behaves similar to $\rho_{xz}(H)$ and $S_{xz}(H)$ since both $\rho_{xx}$ and $S_{zz}$ barely depend on magnetic field. In Figure 3D we plot the temperature dependence of the anomalous component $\alpha_{xz}^A$ extracted at 2 T. $\alpha_{xz}^A$ increases with increasing temperature and tends to saturate at high temperature, reaching a value of 4.6 A m$^{-1}$ K$^{-1}$ at 300 K. Such behavior is significantly different from that of conventional ferromagnetic metals, where $\alpha_{xz}^A$ increases linearly with decreasing temperature and follows the temperature dependence of magnetization[42].

To better understand the origin of the AHE and ANE features observed in Fe3Ge, we performed theoretical calculations to obtain the electronic structure, Berry curvature, anomalous Hall conductivity, and anomalous thermoelectric conductivity. Figure 3E shows the calculated electronic structure and the Berry curvature distribution. Without taking into account the spin-orbit coupling, multiple band crossings occur near the Fermi level, three of which are positioned exactly at the Fermi level along the Γ–M, Γ–K, and Γ–A directions. Spin-orbit coupling induces small gaps at these crossings, leading to significant Berry curvature, as shown in the Berry curvature distribution. Notably, the Berry curvature at the two edges of each gap exhibits opposite signs. This substantial Berry curvature gives rise to a large anomalous Hall conductivity, which strongly depends on the position of the Fermi Level as illustrated in the inset of Figure 2F. While the experimental AHE $|\sigma_{zx}^A|$ at 60 K is slightly smaller than the theoretical AHE at the charge neutral point $E_0$: $\sigma_{zx}^A \sim 400$ ($\Omega^{-1}$cm$^{-1}$), it is closer to the calculated value at –15 meV or 35 meV relative to $E_0$.

The large Berry curvature near the Fermi level also enhances the ANE. Figure 3F shows the temperature dependence of $\alpha_{xz}^A$ at several representative chemical potentials, which was calculated using Kubo formula, $\alpha_{xy}^{int}(T) = -\frac{e}{\hbar}\int d\zeta \frac{\partial f(\zeta-\mu)}{\partial \zeta} \frac{\zeta-\mu}{T} \int_{BZ} \frac{d\vec{k}}{(2\pi)^3} \sum_{\epsilon_n < \zeta} \Omega_{xy}^z(\vec{k})$, wherein $f(\zeta - \mu) = \left(e^{\frac{\zeta-\mu}{k_BT}} + 1\right)$ is the Fermi-Dirac distribution function and $\mu$ is the chemical potential. As it shows, $\alpha_{xz}^A$ increases with increasing the temperature when the chemical potential is below about –50 meV (relative to the charge neutral point). The overall magnitude of $\alpha_{xz}^A$ is also comparable with the experimental values, affirming that the large $\alpha_{xz}^A$ is intrinsic and arises from the Berry curvature of the gapped bands near the crossings as shown in Figure 3E. Figure 3G presents the temperature dependence of $|\alpha_{xz}^A/\sigma_{xz}^A|$ ratio of Fe$_3$Ge and its comparison with other magnetic materials. While $\sigma_{xz}^A$ is related to the Berry curvature of electronic bands over the whole Fermi sea and $\alpha_{xz}^A$ is sensitive to the Berry curvature of the electronic structure near the Fermi level, it was reported that this ratio shows a universal feature: approaching to zero at the low temperature limit but to $\frac{k_B}{e}$ = 86 (µV/K) at high temperature[41]. Indeed, as shown in Figure 3G, this ratio of Fe$_3$Ge increases with temperature and approaches to $\frac{k_B}{e}$ around room temperature, a feature similar to that observed in other topological materials such as Mn$_3$Ge and Co$_2$MnGa [41]. In Figure 3H we plot the maximum $\alpha_{xz}^A$ value of Fe$_3$Ge in comparison with some conventional ferromagnets and well-known topological ferromagnets[20,23,40,43-52], highlighting the large transverse thermoelectric effect in Fe$_3$Ge. This, together with the high magnetic ordering temperature, renders Fe$_3$Ge an excellent candidate with great potential for practical thermoelectric applications based on Nernst effect.

**Topological Hall effect and topological Nernst effect** In addition to the AHE and ANE, we also observe topological Hall effect (THE) and topological Nernst effect (TNE) in Fe$_3$Ge single crystals. In some systems with nontrivial spin structure, the THE may emerge together with the ordinary Hall effect and ANE within a certain magnetic field-temperature phase region[53,54]. That is, $\rho_{xz} = \rho_{xz}^O + \rho_{xz}^A + \rho_{xz}^T = R_0\mu_0 H + S_A\rho_{xx}\rho_{zz}M + \rho_{xz}^T$, where the $R_0$, $S_A\rho_{xx}\rho_{zz}$, and $\rho_{xz}^T$ are the ordinary Hall coefficient, anomalous Hall coefficient, and the topological Hall resistivity, respectively. Thus, the $\rho_{xz}^T$ can be calculated by subtracting both the ordinary and anomalous Hall components from the total Hall resistivity $\rho_{xz}$. At a fixed temperature, since both $\rho_{xx}$ and $\rho_{zz}$ are weakly field-dependent, $\rho_{xz}^A$ effectively depends linearly on $M$. In Figure 4A, we overplot the $\rho_{xz}(H)$ and the scaled $M(H)$ curves measured at 260 K with the magnetic field applied within the *ab* plane. Note that the magnetization and transport measurements were carried out on the exact same piece of single crystal sample to eliminate potential effects of different demagnetization fields due to different sample geometries. The difference in the saturation fields of $M(H)$ and $\rho_{xz}(H)$ data, which was also observed in polycrystalline Fe$_3$Ge [31], implies the emergence of topological Hall effect in this system. Since the ordinary Hall component is much smaller than the other two components, the difference between total $\rho_{xz}$ and $\rho_{xz}^A$ (i.e., the scaled $M$ term) gives $\rho_{xz}^T$ which is represented by the green curves shown in Figure 4A (also see Fig. S2 and the related discussion in the Supplemental Information[32]). Following a similar approach, we can extract the TNE contribution from the total Nernst effect coefficient $S_{xz}$ with $S_{xz} = S_0 + S_{xz}^A + S_{xz}^T$, where $S_0$, $S_{xz}^A$, and $S_{xz}^T$ denote the ordinary (~ $H$), anomalous (~ $M$), and topological Nernst components, respectively. The green curve shown in Figure 4B represents the thus-extracted TNE $S_{xz}^T$. It is seen that both THE and TNE exhibit a maximum near 0.6 T, where the magnetization reaches ~ 90% of the saturated value (see Figure S3 and the related discussion in the Supplemental Information), implying the persistence of a residual non-collinear component. Upon further increasing the magnetic field, both THE and TNE signals decrease and approach to zero when the magnetization saturates at ~ 1.4 T. This suggests that the THE and TNE may arise from the magnetic field induced non-coplanar spin structure in Fe$_3$Ge, which gives non-zero spin chirality and thus non-zero Berry curvature. Figure 4C and 4D present the thus-extracted $\rho_{xz}^T$ and $S_{xz}^T$ at various temperatures, respectively, and Figure 4E and 4F show the corresponding magnetic field-temperature contour maps. Both $\rho_{xz}^T$ and $S_{xz}^T$ increase with temperature, reaching a value of 0.9 $\mu\Omega$ cm and 1.4 $\mu$V/K at 320 K, respectively, which are comparable to or surpass values of some related compounds, such as Fe$_3$Sn$_2$ [49], ScMn$_6$Sn$_6$ [55], MnGe [56], Gd$_2$PdSi$_3$ [57], EuCd$_2$As$_2$ [58], Nd$_3$Ru$_4$Al$_{12}$ [59], YMn$_6$Sn$_6$ [60,61], ErMn$_6$Sn$_6$ [62,63], Mn$_5$Si$_3$ [64], and Mn$_3$Ga [65]. Note that the THE was previously reported in Ref. [31] on the studies of Fe$_3$Ge polycrystalline samples but not in Ref. [30] and Ref. [37] on both polycrystalline and single crystalline samples.

The observation of the THE in Fe$_3$Ge, which has a collinear ferromagnetic structure at zero field, is intriguing. Generally, the emergence of the THE stems from non-zero static scalar spin chirality (SSC) in systems with non-coplanar spin texture (e.g, the formation of Skyrmion lattice) which induces a nontrivial real-space Berry phase and the associated fictitious magnetic field[53,54]. More recently, a fluctuation-driven mechanism has been proposed to generate a finite SSC and account for the observation of THE in systems without static SSC (such as YMn$_6$Sn$_6$ [60] and Fe$_3$Ga$_4$ [66,67]), which involves chiral spin fluctuations. Although these materials exhibit a helical spin spiral ground state, it transitions—under an in-plane magnetic field—into a transverse conical spiral (TCS) phase, where the spiral propagates out-of-plane (along the c-axis) while spins rotate in the ab-plane. This TCS structure still lacks static scalar spin chirality yet enables the population of thermally excited chiral magnons with a preferred handedness. These magnons become unequally populated at finite temperature, dynamically generating a nonzero scalar spin chirality and giving rise to the THE[60,62,67]. As discussed previously, Fe$_3$Ge exhibits spin reorientation from the out-of-plane alignment above $T_{SR} \sim 385$ K to the in-plane alignment at lower temperatures, with zero-field neutron diffraction measurements revealing mixed in-plane and out-of-plane moment components near $T_{SR}$ (Fig. 1E and Ref.[33]). This suggests that as the temperature approaches T$_{SR}$, the system's reduced anisotropy barriers may enable the formation of non-coplanar spin configurations in the presence of magnetic field, supported by recent first-principles predictions[31]. Very recent Lorentz Transmission Electron Microscopy measurements conducted at room temperature with in-plane applied magnetic fields reveal no evidence of Skyrmion spin textures in Fe$_3$Ge.[68] It is likely that the field-induced non-coplanar spin structure together with the chiral spin fluctuations gives rise to the observed THE in this system. Thus, future neutron scattering studies in the presence of magnetic field to probe the evolution of spin structure are highly desirable. It is worth pointing out that the THE and TNE peak position barely depends on the measurement temperature up to 320 K, as shown in Fig. 4(E,F). Although the underlying mechanism is still elusive, this feature is similar to the THE reported in YMn$_6$Sn$_6$ and ErMn$_6$Sn$_6$ in which the chiral spin fluctuations are found to be the driving mechanism of the observed THE [60,62]. On the other hand, for the chiral fluctuation driven topological Hall effect, it was shown that $\rho^T$ at a constant field is proportional to temperature (i.e, $\rho^T \sim T$), as shown in Figure S4(c) for YMn$_6$Sn$_6$ [60]. The nearly linear temperature dependence of $\rho^T_{xz}$ and $S^T_{xz}$ of Fe$_3$Ge at 0.6 T shown in Figure S4 (a,b) implies that the topological Hall effect and topological Nernst effect in this system may be associated with chiral spin fluctuation driven non-zero scalar chirality as well.

In summary, we show that Fe$_3$Ge exhibits large anomalous Hall effect and anomalous Nernst effect with the anomalous Nernst conductivity reaching 4.6 A m$^{-1}$ K$^{-1}$. Our first-principles calculations qualitatively reproduce the observed anomalous transverse transport results, highlighting the critical contribution from the Berry curvature of the massive Dirac gaps in the momentum space. In addition, both topological Hall effect and topological Nernst effect are also observed, with the topological Nernst coefficient increasing when the temperature approaches the spin-orientation phase transition, which presumably arises from the Berry phase associated with the field-induced non-zero scalar spin chirality. These results highlight the synergic effects of the Berry phases in both momentum space and real space of Fe$_3$Ge. This, in conjunction with its high magnetic ordering temperature, places this material a promising candidate for potential thermoelectric applications based on Nernst effect.

**Acknowledgments** S. G., O.E., and X.K. acknowledge the financial support by the U.S. Department of Energy, Office of Science, Office of Basic Energy Sciences, Materials Sciences and Engineering Division under Grant No. DE-SC0019259. The thermoelectric transport measurements were supported by National Science Foundation (DMR-2219046). M.X. and W.X. acknowledge the financial support by the U.S. Department of Energy, Office of Science, Office of Basic Energy Sciences, Materials Sciences and Engineering Division S under Contract DE-SC0023648. P. P. Zhang acknowledges the financial support from the U.S. Department of Energy, Office of Basic Energy Sciences, Division of Materials Sciences and Engineering under Award Number DE-SC0019120. Neutron diffraction part of this research used resources at the High Flux Isotope Reactor, a DOE Office of Science User Facility operated by the Oak Ridge National Laboratory. The beam time was allocated to DEMAND on proposal number IPTS-32305.1. C.X. was partially supported by the Start-up funds at Michigan State University and National Natural Science Foundation of China (Grants No. 12304071).

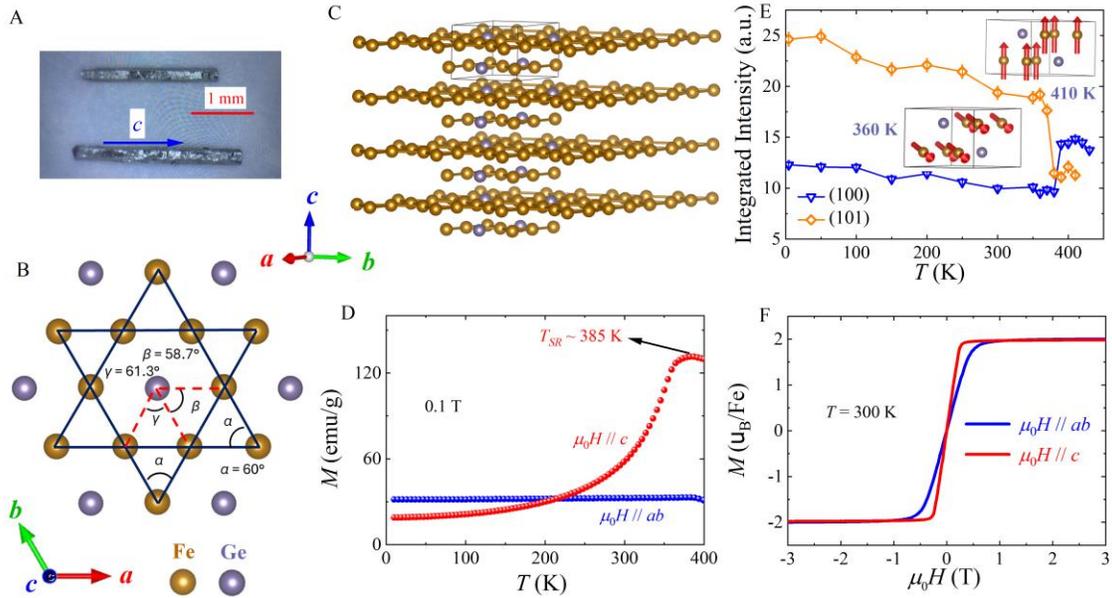

**Figure 1: Crystal structure and magnetic properties.** (A) An optical image of the as-grown Fe$_3$Ge single crystals. (B, C) Schematic view of the crystal structure of Fe$_3$Ge from two axes perspective, respectively. (D) Temperature dependence of magnetic susceptibility measured under 0.1 T. (E) The temperature dependence of integrated neutron scattering intensity of (100) and (101), the insets show the schematic diagram of spin configuration is indicated by red arrows at 360 K and 400 K. (F) The field dependence of magnetic susceptibility measured at 300 K.

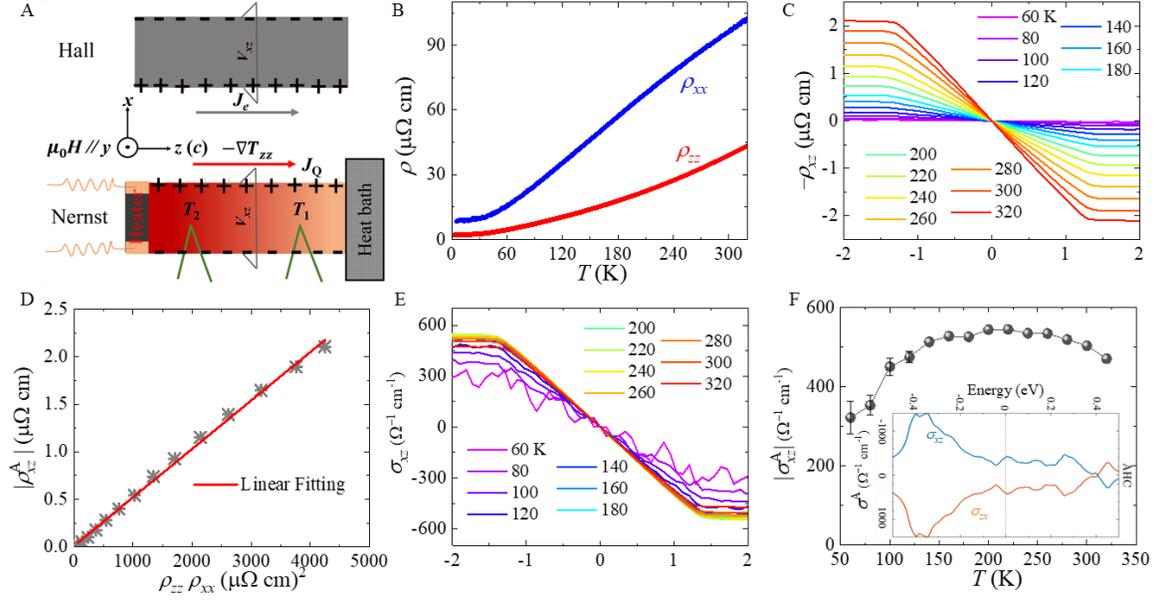

**Figure 2: Anomalous Hall effect.** (A) The schematic diagram of Hall effect and Nernst effect measurements. Note that $J_Q \parallel c$, $H \perp c$. (B) The temperature dependence of longitudinal resistivity $\rho_{xx}$ and $\rho_{zz}$ (C) The Hall resistivity $\rho_{zx}$ at different temperatures. (D) Plot of $\rho_{zx}^A$ vs $\rho_{xx}\rho_{zz}$, the red solid line is the linear fitting. (E) The Hall conductivity $\sigma_{xz}$ at different temperatures. (F) The temperature dependence of experimental anomalous Hall conductivity $|\sigma_{xz}^A|$, the inset shows the theoretically calculated intrinsic $\sigma_{xz}^A$ as a function of the chemical potential, where the magnetization is along one of the in-plane lattice vectors.

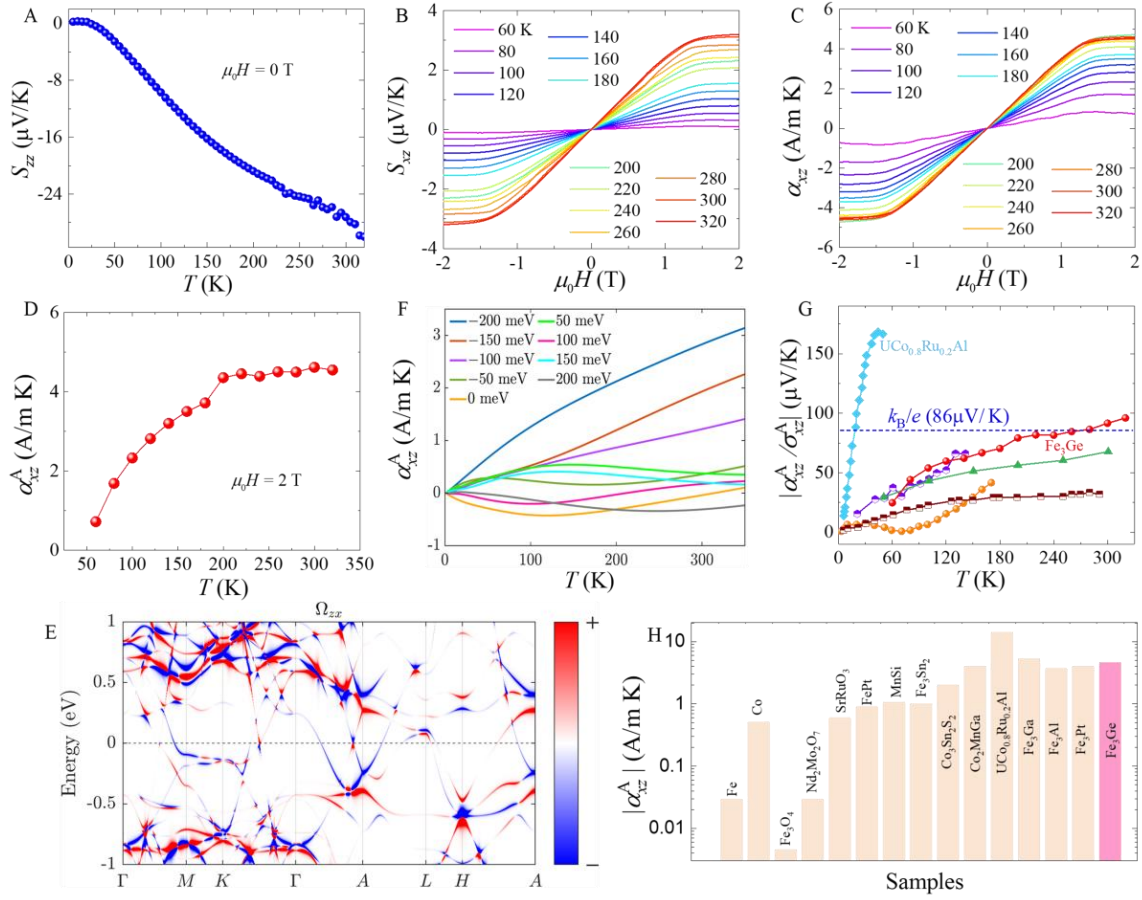

**Figure 3: Anomalous transverse thermoelectric effect.** (A) The temperature dependence of Seebeck coefficient $S_{zz}$ under 0 magnetic field. (B) The Nernst signals as a function of field at different temperatures. (C) The magnetic field dependence of the calculated transverse thermoelectric conductivity $\alpha_{xz}$ using the five experimentally measured components $\rho_{xx}$, $\rho_{zz}$, $\rho_{xz}$, $S_{zz}$, $S_{xz}$ at various temperatures. (D) The temperature dependence $\alpha_{xz}^A$ extracted from panel C at 2 T. (E) The electronic band structure and the projected Berry phase for bands near the Fermi energy. (F) The temperature dependence of the theoretically calculated $\alpha_{xz}^A$ at representative chemical potentials. (G) The ratio $|\alpha_{xz}^A/\sigma_{xz}^A|$ of Fe$_3$Ge as a function of temperature in comparison with other magnets[23]. (H) The value of $|\alpha_{xz}^A|$ of Fe$_3$Ge, compared with some well-known conventional ferromagnets and topological ferromagnets (Fe[50], Co[47], Fe$_3$O$_4$[40], Nd$_2$Mo$_2$O$_7$[45], SrRuO$_3$[43], FePt[48], MnSi[46], Fe$_3$Sn$_2$[49], Co$_3$Sn$_2$S$_2$[20], Co$_2$MnGa[44], UCo$_{0.8}$Ru$_{0.2}$Al[23], Fe$_3$Ga[51], Fe$_3$Al[51], and Fe$_3$Pt[52]).

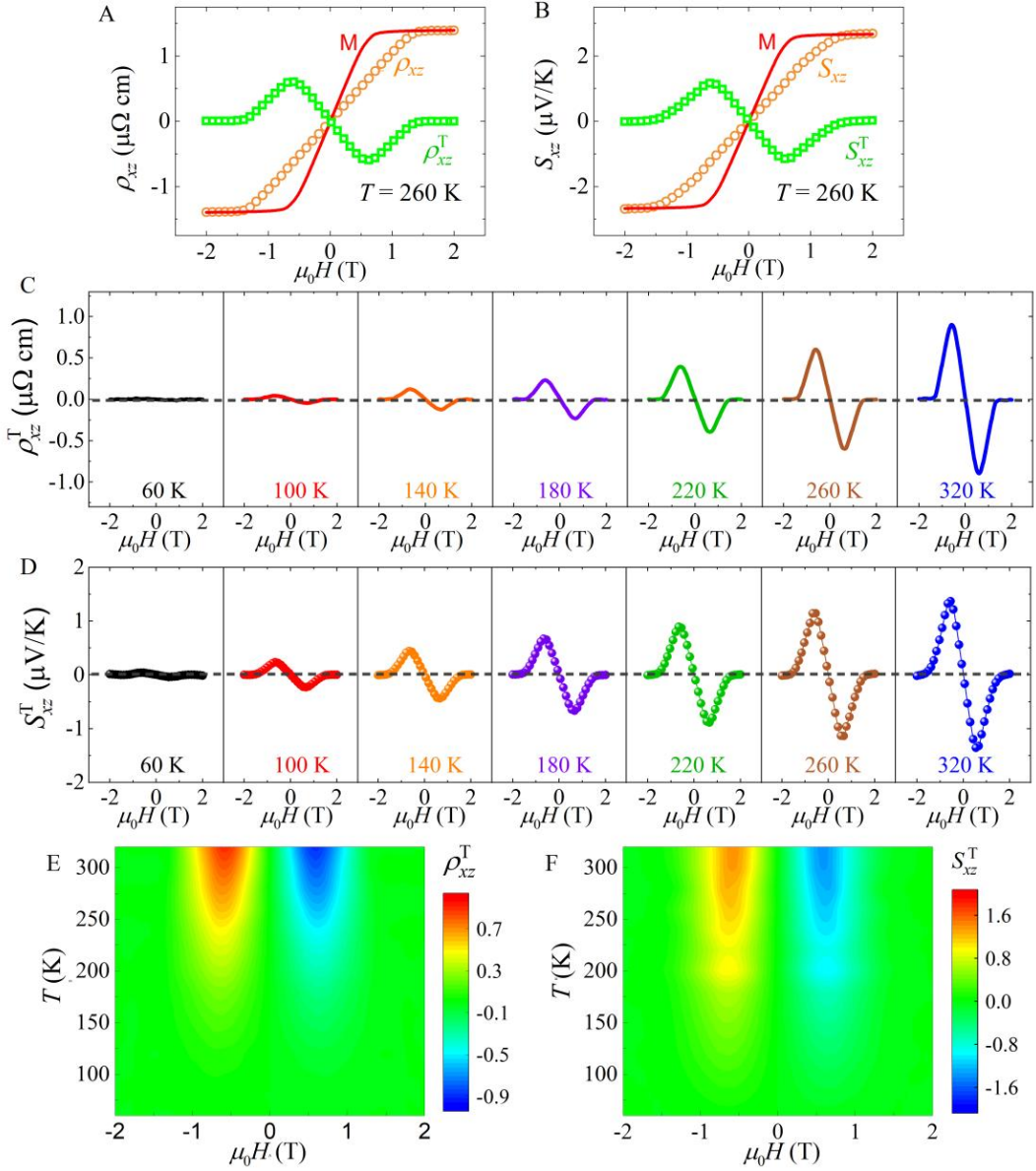

**Figure 4: Topological Hall effect and topological Nernst effect.** (A) Magnetic field dependence of total Hall resistivity $\rho_{xz}$ and the scaled magnetization at 260 K. Green curve stands for the extracted topological Hall $\rho_{xz}^T$ described in the main text; (B) Magnetic field dependence of total Nernst effect coefficient $S_{xz}$ and the scaled magnetization at 260 K. Green curve stands for the extracted topological Nernst effect coefficient $S_{xz}^T$; (C) Magnetic field dependence of topological Hall $\rho_{xz}^T$ at some selected temperatures. (D) The topological Nernst effect signal $S_{xz}^T$ as a function of magnetic field at some selected temperatures. (E) The $T$-$\mu_0 H$ contour map of $\rho_{xz}^T$. (F) The $T$-$\mu_0 H$ contour map of $S_{xz}^T$.